# Electrical detection of the spin-flop and room-temperature magnetic ordering in van der Waals CrPS$_4$/(Pt, Pd) heterostructures

Rui Wu[1,2,3], Andrew Ross[1,4], Shilei Ding[1,5], Yuxuan Peng[5], Fangge He[1], Yi Ren[3], Romain Lebrun[1,4], Yong Wu[1,6], Zhen Wang[1,7], Jinbo Yang[5], Arne Brataas[2,*], Mathias Kläui[1,2,*]

1. Institute of Physics, Johannes Gutenberg-University Mainz, Staudingerweg 7, Mainz 55128, Germany
2. Center for Quantum Spintronics, Norwegian University of Science and Technology, Trondheim 7491, Norway
3. Beijing Academy of Quantum Information Sciences, Beijing, 100193, P. R. China
4. Unité Mixte de Physique, CNRS, Thales, Université Paris-Saclay, Palaiseau, 91767, France.
5. State Key Laboratory for Mesoscopic Physics, School of Physics, Peking University, Beijing 100871, China
6. School of Materials Science and Engineering, University of Science and Technology Beijing, Beijing 100083, China
7. Department of Applied Physics, Chang'an University, Xi'an 710064, China
*Corresponding authors: arne.brataas@ntnu.no; klaeui@uni-mainz.de

**We study magneto-transport in heterostructures composed of the van der Waals antiferromagnet CrPS$_4$ and the heavy metals Pt and Pd. The transverse resistance ($R_{xy}$) signal reveals the spin-flop transition of CrPS$_4$ and a strongly enhanced magnetic ordering temperature (>300 K), which might originate from a strong spin-orbit coupling at the interface. While CrPS$_4$/Pt devices allow for easy detection of the spin-flop transition, CrPS$_4$/Pd devices show a more substantial enhancement in magnetic ordering temperature and exhibit a topological Hall effect signal, possibly related to chiral spin structures at the interface. The longitudinal magnetoresistance ($R_{xx}$) results from a combination of spin-Hall magnetoresistance and the negative magnetoresistance that can be explained by a field-induced change of the electronic band structure of CrPS$_4$.**

The coupling between spin and orbital degrees of freedom in materials, especially in heavy metals such as Pt, Pd and W, and in topological insulators, has led to new directions in spintronics, such as spin-orbitronics[1-4]. In metals with significant spin-orbit coupling, the spin Hall effect (SHE)[5] generates a spin current from a charge current that in turn interacts with adjacent magnetic materials[6]. Spin-orbit coupling has opened a new route to detect and manipulate the magnetic properties of magnetic materials. For example, the spin-Hall magnetoresistance (SMR)[6], originating from the interaction of the spin currents generated by the heavy metals and the magnetic layer that can be described as a nonequilibrium proximity effect, is a useful tool to probe the magnetic moments of the magnetic materials. The SMR is especially important in magnetic insulators, where a direct measurement with electric currents flowing in the magnetic layer is not possible.

Van der Waals (vdW) magnets show extraordinary prospects for spintronic devices with reduced dimensionality, high flexibility, and tunability[7-13]. The $CrX_3$ (X = I, Cl, Br) series is an example of typical vdW magnets. The bulk materials have A-type antiferromagnetic ordering so that the magnetic moments within each layer are parallel as in ferromagnets, while the inter-layer coupling is antiferromagnetic[14]. Significantly, the properties are maintained even at the monolayer scale despite the Mermin-Wagner theorem[15] because of the sizeable magnetic anisotropy. Furthermore, the antiferromagnetic coupling between layers causes the magnetic properties to differ for odd and even numbers of stacking layers[14, 16]. An applied electric field can control the magnetism of a double-layer vdW magnet by controlling the relative electron occupancy of the two sublattices and the inter-layer exchange coupling[17]. Additionally, mechanical pressure can control the inter-layer coupling to switch the configuration from antiferromagnetic to ferromagnetic[16]. At present, spintronic devices based on this series of vdW magnets are of considerable interest, and there have been observations of a large tunneling magnetoresistance in the MTJs[18, 19]. However, the $CrX_3$ class of materials is unstable in both air and light, limiting their easy application in practical devices.

A more robust alternative is the thiophosphate (CrPS$_4$) that also belongs to the A-type antiferromagnet family and has a layered structure with vdW interlayer interactions, where spins within each monolayer are aligned ferromagnetically out-of-plane, as schematically shown in Figure 1(a)[20, 21]. Both density functional theory (DFT) calculations [22] and experimental results[23] have indicated that single-layer CrPS$_4$ is a 2-dimensional ferromagnet. Compared with some members of the CrX$_3$ series, CrPS$_4$ is air-stable and has a sizeable Néel temperature ($T_N$=36 K)[24]. A $T_C$ = 25 K is observed in single-layer CrPS$_4$, which is smaller than the bulk $T_N$[23]. In addition, CrPS$_4$ exhibits a spin-flop (SF) transition, where the Néel vector **n** is rotated by the external magnetic field **H** from **n**//**H** to **n**⊥**H** configurations. The SF can minimize the effective anisotropic energy that the spin experiences and allows for tuning of the effective anisotropy, for instance crucial for realizing devices based on the long-distance spin transport in antiferromagnetic materials in the superfluid or diffusion regimes[25-29]. Moreover, the SF field is $H_{sf}$ = 0.9 T at 5 K for bulk CrPS$_4$[24], which is relatively accessible compared to other antiferromagnetic systems with a $H_{sf}$ of several Tesla or tens of Tesla[24, 30, 31]. For application, electrical detection of the antiferromagnetic order is key. However, the electrical readout of CrPS$_4$ has not been reported to date, calling for the identification of magnetoresistance effects that lend themselves to electrical detection of the magnetic order.

In this paper, we report measurements of magneto-transport in CrPS$_4$/heavy metal (Pt, Pd) heterostructures. The Hall effect measurement reveals the SF transition，indicating a spin transparent vdW/heavy metal interface. Moreover, we measure a persistence of the AHE in the heterostructures at temperatures above the $T_N$ of the bulk CrPS$_4$, and up to 300 K. Especially, we report a topological Hall effect (THE) in the high-temperature range up to 240 K in the device with Pd but not in the device with Pt, that can be explained by a smaller Dzyaloshinskii–Moriya interaction (DMI) at the vdW/heavy metal interface in the Pt based devices possibly related to the different Pt and Pd work functions. Furthermore, for a Pd based heterostructure, the field dependence of the longitudinal resistance shows a strong ferromagnetic-like signal at temperatures up to 300 K as well as a strong negative magnetoresistance at temperatures below $T_N$, possibly related to the electronic band structure change in CrPS$_4$ induced by the external magnetic field. The angular dependence of longitudinal resistance confirms the ferromagnetic ordering above the bulk $T_N$ of CrPS$_4$ and shows a signature of the SF at temperatures below $T_N$. We ascribe the enhancement of the ordering temperature to the strong spin-orbit interaction supplied by the heavy metal layer, which makes the top layer of CrPS$_4$ in contact with heavy metals stay ferromagnetic at elevated temperatures.

The CrPS$_4$ flakes are mechanically exfoliated from CrPS$_4$ single crystals with adhesive tape and transferred to an α-Al$_2$O$_3$(0001) substrate. Raman spectroscopy shown in Figure S1 indicates the good quality of the CrPS$_4$ flakes. An atomic force microscopy image of a typical flake is shown in Figure 1(b). The profile across the flake edge indicates that its thickness is about 40 nm. To maintain a clean interface, we complete the entire process in a glove box directly connected to a Molecular-beam epitaxy (MBE) deposition system. Next, we transfer the substrates with CrPS$_4$ flakes to the MBE without exposure to air. During the metal deposition process, the substrate is cooled to about 90 K by liquid nitrogen to reduce any damage to the CrPS$_4$ surface during the deposition process. The deposition rate is 1 nm/min. A reference CrPS$_4$/Pd sample with Pd deposited using magnetron sputtering at room temperature also without exposure to air does not significantly alter the transport properties. Thus, we prepared the CrPS$_4$/Pt sample by magnetron sputtering, exfoliating the flakes in a load-lock chamber connected to the main sputtering chamber in a vacuum of about $5\times10^{-6}$ mbar. The thickness of the heavy metal layers is about 10 nm. Finally, we obtained Hall devices illustrated in Figure 1(c) by photolithography and ion beam etching. To obtain better conductivity at the edge of the sheet, we deposited 60 nm thick Au contacts to cover the flake's edge, as shown in Figure 1(d).

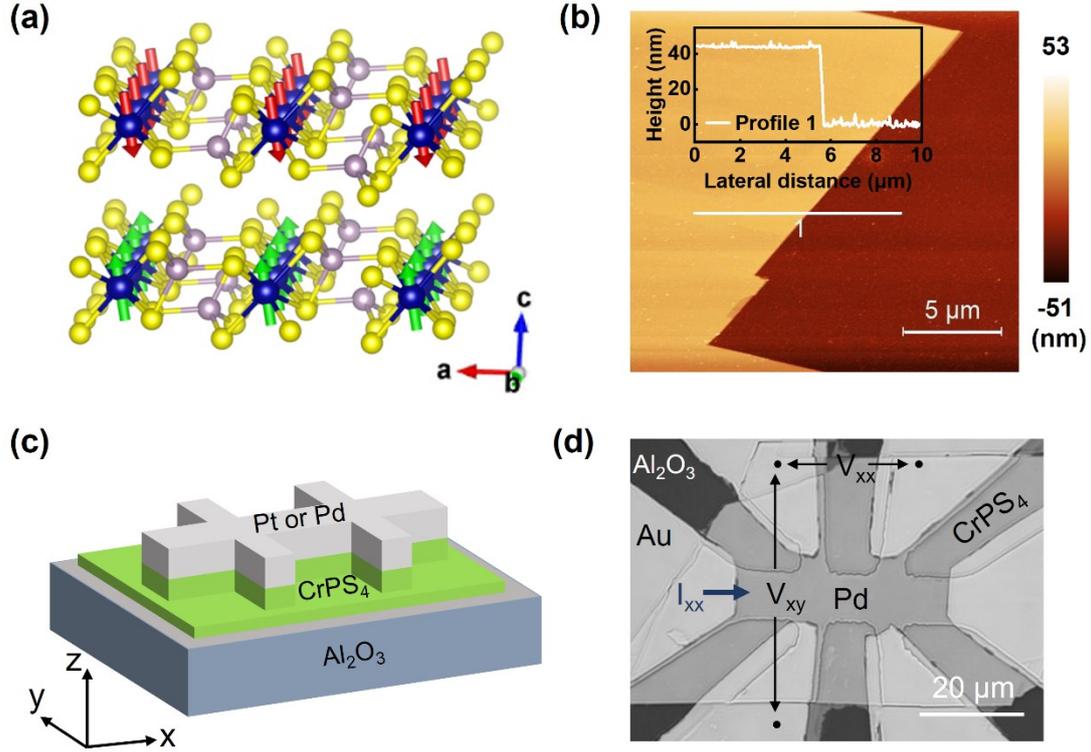

**Figure 1.** (a) Crystalline and magnetic structures of CrPS$_4$. (b) Atomic force microscopy image of a CrPS$_4$ flake with a thickness of about 40 nm. The inset shows the profile across the flake edge. (c) Schematic of the CrPS$_4$/(Pt, Pd) device. (d) Optical microscope image of a CrPS$_4$/Pd device.

We first studied the transport in the Hall geometry of the CrPS$_4$/(Pt, Pd) devices. With the magnetic field applied in the out-of-plane direction (z-axis) and a constant current $I_{xx}$=1 mA applied along the longitudinal channel, the $V_{xy}$ was measured while varying the magnetic field and external temperature and shown in Figure 1(d). Considering that the width of the current channel in the Hall bar is about 9 μm, the current density **J** is about 1.1×10$^6$ A/cm$^2$. The transverse resistance is defined as $R_{xy} = V_{xy}/I_{xx}$. We have subtracted a linear background for the $R_{xy}$ signal due to the normal Hall effect of Pt and Pd, as shown in Figure S2.

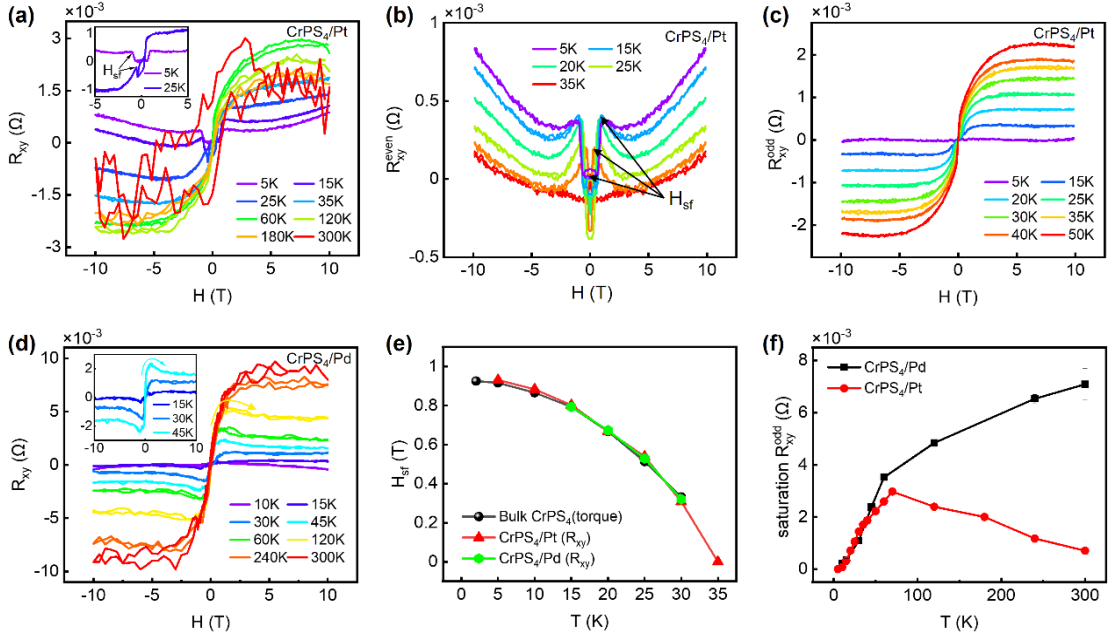

**Figure 2.** (a) $R_{xy}$ signal of CrPS$_4$/Pt device measured at different temperatures, and the inset shows the zoomed-in curves obtained at 5 K and 25 K. (b) The even component ($R_{xy}^{even}$) odd component ($R_{xy}^{odd}$) of $R_{xy}$ of the CrPS$_4$/Pt device are presented. (d) The $R_{xy}$ of CrPS$_4$/Pd device measured at different temperatures, and the inset shows the zoomed-in curves obtained at 15 K, 30 K, and 45 K. (e) The SF fields obtained from CrPS$_4$/Pt (red triangle) and CrPS$_4$/Pd (blue diamond) devices in comparison with values from the bulk CrPS$_4$ (black sphere, data adapted from Ref. [17]). (f) The comparison of the saturation of the $R_{xy}^{odd}$ versus temperature plots for the CrPS$_4$/Pd (black square) and CrPS$_4$/Pt (red circle) devices.

As shown in Figure 2(a), there are two central features in the $R_{xy}$ curves of the CrPS$_4$/Pt device. Firstly, at temperatures $T < T_N$, the $R_{xy}$ curves show significant jumps at a field $H_{sf}$, and they are asymmetric as a function of magnetic field, see the inset of Figure 2(a). This feature has, to the best of our knowledge, not been seen in other magnetic systems and we next check if it is related to the SF of the CrPS$_4$. In order to separate the possible SF signal from the aforementioned AHE, which should be symmetric, we separate $R_{xy}$ into an even component and an odd component, i.e. $R_{xy}(H) = R_{xy}^{even}(H) + R_{xy}^{odd}(H)$, where

$$R_{xy}^{odd}(H) = \frac{R_{xy}(H) - R_{xy}(-H)}{2},$$

$$R_{xy}^{even}(H) = \frac{R_{xy}(H) + R_{xy}(-H)}{2}.$$

Figure 2(b) and 2(c) show $R_{xy}^{even}(H)$ and $R_{xy}^{odd}(H)$ at different temperatures. It is found that the jumps in the $R_{xy}$ curves are only observed in $R_{xy}^{even}$ signal at temperature below $T_N$. The slow increase of $R_{xy}^{even}$ at higher fields is likely caused by a small contribution from the longitudinal resistance $R_{xx}$ due to the small (inevitable) offset of the $V_{xy}$ contacts[32]. Finally, we now plot the positions of the jumps as a function of the temperature in Figure 2(e). The position of the jumps agrees extremely well with the SF field $H_{sf}$ in the bulk CrPS$_4$ obtained from torque measurements[21], as compared in Figure 2(f). When the temperature is close to $T_N$, the jumps tend to overlap with each other (see the results for T=35 K in Figure 2(b)). Thus, this shows that we can detect the spin-flop electrically in this system as a key step for electrical read-out.

Secondly, as shown in Figure 2(a), the amplitude of $R_{xy}$ shows a non-monotonic temperature dependance. This feature is majorly reflected in the $R_{xy}^{odd}$ signal as shown in Figure 2(c). The saturation $R_{xy}^{odd}$ monotonically increases with increasing temperature when T < 70 K and then decreases when T > 70 K, as shown in Figure 2(f). It is found that a non-zero $R_{xy}^{odd}$ can be found even at room temperature, far above the bulk $T_N$. This behavior contrasts with the usual situation where $R_{xy}$ due to the anomalous Hall effect (AHE) in ferromagnetic thin films decreases with the increase of the temperature and disappears at the Curie temperature $(T_C)$[33]. It is noted that no AHE observed in the Hall bar of pure heavy metals (Pt or Pd) in our experiment, the observed FM-like signal of the CrPS$_4$/Pt at above $T_N$ is ascribed to the enhanced ordering temperature of the interface CrPS$_4$ layer in contact with Pt and with an enhanced spin-orbit coupling[34]. The enhancement of magnetic ordering temperature has been observed in several vdW/heavy metal heterostructures, including the Cr$_2$Ge$_2$Te$_6$/W and Cr$_2$Ge$_2$Te$_6$/Ta systems, due to the proximity effect from the heavy metals[35]. As we know, the bulk CrPS$_4$ has a layered structure and a weak interlayer exchange coupling. When

$T > T_N$, the bulk CrPS$_4$ becomes paramagnetic while the interface CrPS$_4$ can stay ferromagnetic because the extended 5d orbitals of heavy metals with strong SOC can participate in the reconstructed orbital hybridization in the adjacent CrPS$_4$ layer and promote the intralayer exchange interaction[35]. The different temperature dependence below and above 70 K is from the competition of thermally enhanced charge transfer between Pt and CrPS$_4$, which enhances the magnetic ordering, and the ferromagnetic to paramagnetic transition induced by increasing the temperature.

As shown in the Figure 2(d), the main features of the $R_{xy}$ signal are also observed in the CrPS$_4$/Pd sample. Firstly, when $T < T_N$, the asymmetry and jumps in the $R_{xy}$ curves is also present in CrPS$_4$/Pd device, as shown in the inset of Figure 2(d) and Figure S3(a). The $R_{xy}^{even}$ and $R_{xy}^{odd}$ are calculated and given in Figure S3(b) and S3(c), respectively. As shown in Figure S3(b), the jumps in the $R_{xy}$ are clearly visible in the $R_{xy}^{even}(H)$ signal in a narrow temperature range of 15 K-30 K. The position of the jumps again agrees extremely well with the $H_{sf}$ obtained from the torque measurement of the bulk CrPS$_4$[21], as compared in Figure 2(e), confirming that the mechanism for the jumps in the $R_{xy}$ curves is the SF transition in the CrPS$_4$ flake thus showing that again we can realize electrical read-out.

Secondly, the $R_{xy}$ signal of the CrPS$_4$/Pd sample shows a steady increase with increasing temperature, even above the $T_N$ of bulk CrPS$_4$. The difference between these two samples is that the amplitude of $R_{xy}$ of the CrPS$_4$/Pd sample does not reach its maximum even when the temperature is above room temperature as shown in Figure 2(f), while it reaches the maximum at 70 K in the CrPS$_4$/Pt sample. When the temperature is lower than $T_N$ of bulk CrPS$_4$, the two curves almost completely overlap. But when the temperature is higher than $T_N$, they start to diverge, and show a completely different temperature-dependent trend above 70 K. The difference between those two heterostructures can be possibly ascribed to the different electronic properties of the two heavy metals, e.g., the larger work function of Pt ($\Phi_{Pt,vac} = 5.65\ eV$) compared to Pd ($\Phi_{Pd,vac} = 5.12\ eV$)[36] can make the charge transfer process at the CrPS$_4$/Pd interface much easier than that at the CrPS$_4$/Pt interface and give a more prominent $T_C$ enhancement in CrPS$_4$.

Thirdly, as shown in Figure S3(a) and the inset in Figure 2(d), we observe a non-monotonic field dependence in some $R_{xy}$ curves of the CrPS$_4$/Pd device, i.e., when the magnetic field is larger than a particular value, the Hall coefficient $R_{xy}$ drops anomalously and presents a peak in the curve. This behavior becomes visible at 25 K and stronger between 60 K and 120 K before vanishing at 180 K (see the inset in Figure S3(c)). It is reminiscent of signals resulting from a topological Hall effect (THE) [37-39] and usually associated with topological chiral magnetic structures, such as skyrmions and chiral domain walls[40, 41]. These are stabilized by interfaces with DMI, which can be introduced in the ultrathin ferromagnetic layer adjacent to a heavy metal layer due to strong spin-orbit coupling and broken inversion symmetry[41]. The thermal fluctuations have been demonstrated to play a key role in the formation of topological spin textures in Cr$_2$O$_3$/Pt heterostructure, where a THE was observed at 345 K, higher than the T$_N$ of Cr$_2$O$_3$ (~307 K)[42]. The THE in the CrPS$_4$/Pd device at above T$_N$ could result from the same mechanisms. However, the THE is not observed in the CrPS$_4$/Pt device, as shown in Figure 2(a) and 2(b). This absence might arise from different DMI strength at the interfaces in two systems since the DMI is also correlated with the work functions of the heavy metals[43].

Although the SF transition in the $R_{xy}$ signal has been reported in the CoO(001)/Pt system[44], some other systems also show SF transition in the $R_{xx}$ signal, such as CrI$_3$/Pt [45] and α-Fe$_2$O$_3$/Pt[46] systems. Different spin dynamics could yield different behaviors in SMR during the SF transition, considering that the transverse SMR signal is proportional to the in-plane Néel vector components n$_x$·n$_y$, while the longitudinal signal is proportional to n$_y$·n$_y$[6, 47], according to the geometry shown in Figure 2(c).

We next study the field and angular dependance of longitudinal transport properties of the CrPS$_4$/Pd sample to compare this with the transverse signals. For the field dependence, the longitudinal magnetoresistance is defined as $MR_{xx} = \frac{[R_{xx}(H) - R_{xx}(H=0)]}{R_{xx}(H=0)}$, where the longitudinal resistances $R_{xx} = \frac{V_{xx}}{I_{xx}}$, and the current $I_{xx}$ has the same value as in the Hall measurement. Since Pd exhibits a significant Hanle-induced $MR_{xx}$[48], we obtain a signature of CrPS$_4$ in Figure 3 by subtracting the signal of a control Pd device from the signal of the CrPS$_4$/Pd device, i.e., $\delta MR_{xx} = MR_{xx}(\text{CrPS}_4/\text{Pd}) - MR_{xx}(\text{Pd})$ (details in Figure S4).

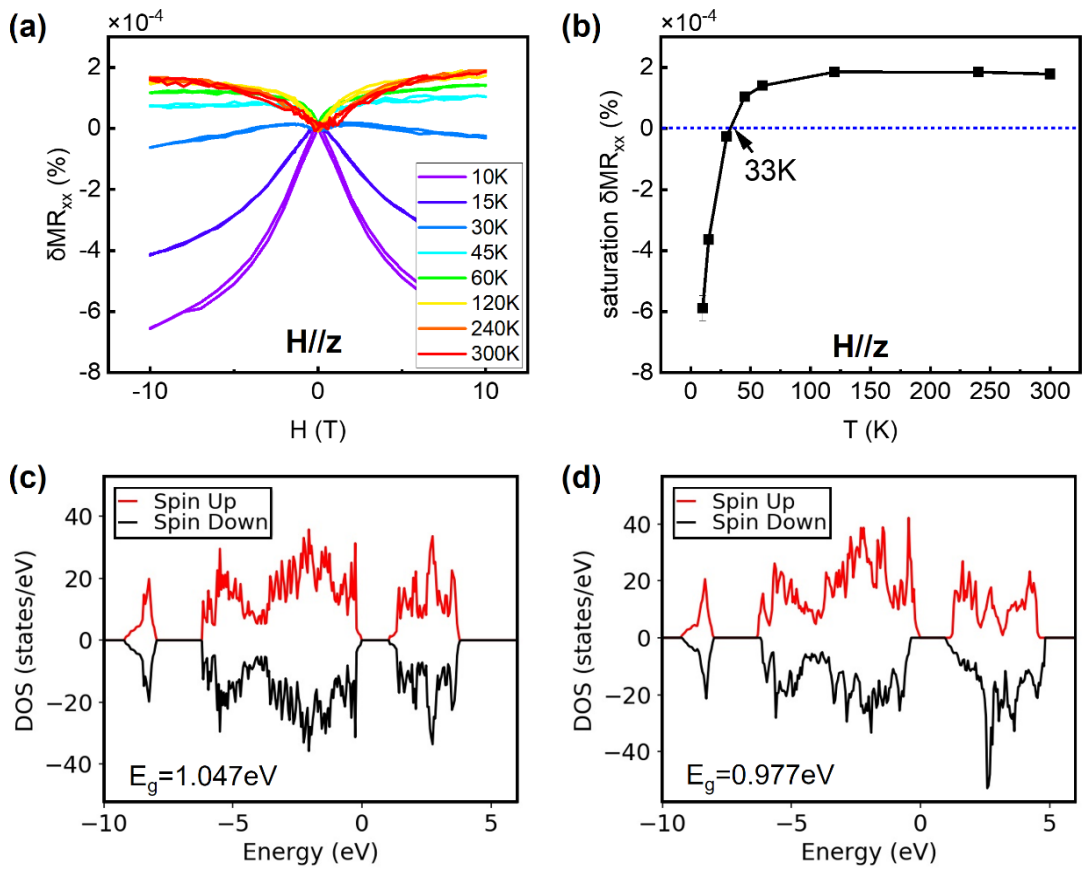

**Figure 3.** (a) The longitudinal magnetoresistance $\delta MR_{xx}$ of the CrPS$_4$/Pd device, as a function of magnetic field applied along z axis. (b) The corresponding saturation $\delta MR_{xx}$ (at 10 T) versus temperature. (c-d) The DFT-calculated DOS of the CrPS$_4$ with antiferromagnetic and ferromagnetic spin structures, respectively.

Figure 3(a) shows the $\delta MR_{xx}$ of the CrPS$_4$/Pd sample measured at different temperatures with the magnetic field applied along the z-axis. The $\delta MR_{xx}$ is negative when the temperature is lower than $T_N$, while it is positive at temperatures above $T_N$ and remains observable even at 300 K. Figure 3(b) plots the nominal saturation $\delta MR_{xx}$ value at the maximum magnetic field (10 T) as a function of temperature, demonstrating a sharp change at low temperature and saturation above 60 K. The sign change is around the Néel temperature $T_N$ of CrPS$_4$. We find a magnetic response at temperature much higher than the $T_N$, which may result from an enhanced magnetic order temperature. The anomalous negative magnetoresistance was found in the measurements with a magnetic field applied along the x and y axes, and all found only below $T_N$ of CrPS$_4$ (details in Figure S5). Thus, this must be related to the electronic structure change in the CrPS$_4$ correlated with a magnetic structure transition from AFM to FM (i.e. the spin-flip process) induced by the magnetic field. This conjecture is supported by DFT calculations (details in the supplementary information), which indicate that the AFM state has a larger bandgap (1.047 eV) than the FM state (0.977 eV), as shown in Figure 3(c) and 3(d), respectively. Furthermore, the $R_{xy}$ signal obtained in the H//x and H//y configurations shows some features related to the spin-flip and magnetic ordering transitions of the bulk CrPS$_4$, respectively (details in Figure S6(a) and S6(b)). This again demonstrates the transparency of the vdW/heavy metal interface in those samples.

It is found that $\delta MR_{xx}$ does not show any SF-related signal in the CrPS$_4$/Pd device, which is different from the previously reported CrI$_3$/Pt device[45], where $R_{xx}$ shows a signal from current induced SF. We conjecture two reasons for this. The first reason is that although CrPS$_4$ has a quite similar magnetic structure to CrI$_3$, the spin dynamics in these two systems are quite different, because here the bulk CrPS$_4$ properties are detected while a surface SF is detected in CrI$_3$ as there is not an intrinsic bulk SF transition in CrI$_3$. Thus, our result represents the first direct electrical readout of the intrinsic SF of a vdW antiferromagnet using transport measurement. Another possible reason is that the SF signal is overwhelmed by the bulk CrPS$_4$ resistance change induced by the magnetic field, and thus cannot be detected, while in the $R_{xy}$ signal, the interference from the bulk CrPS$_4$ signal will be much smaller.

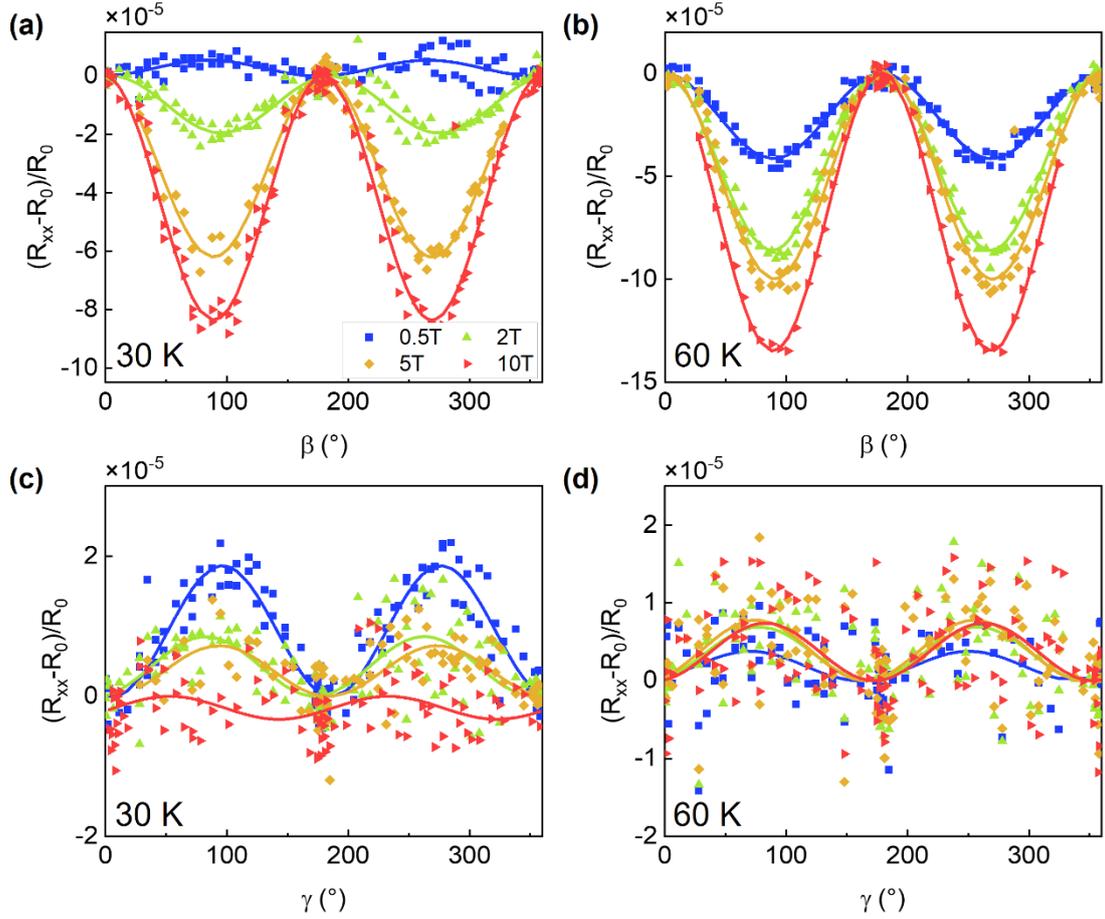

**Figure 4.** Spin-Hall magnetoresistance in the CrPS$_4$/Pd heterostructure. (a) and (b) show the β angle dependence at 30 K and 60 K, respectively. (c) and (d) show the γ angle dependence at 30 K and 60 K, respectively. The scatter indicates the experimental results, and the solid curves indicate the best-fitting of the experimental results.

Finally, we probe the angular variation of the signals by rotating the magnetic field to understand the possible AFM and the FM-like signals and the SMR and the AMR origin better. The angular dependence of the resistance of the CrPS$_4$/Pd device are measured at two different temperatures, 30 K (below $T_N$) and 60 K (above $T_N$). In the SMR scenario, the MR depends on the angle between the nonequilibrium spin accumulation $\sigma$ in the heavy metal layer and the magnetization $M$ of the magnetic layer. In contrast, the AMR depends on the angle between the current $J$ and the magnetization $M$.

Figure 4(a) and 4(b) show the field dependence of $\delta R_{xx}(\beta)$ at 30 K. Here, $\beta$ is the angle between the field and the y axis when the field is rotated in the yz plane, which is zero when the field is parallel to y. Since $M$ is always perpendicular to $J$ in this rotation plane, the magnetoresistance, if detected, should result from SMR only. It is found that the $\delta R_{xx}(\beta)$ shows maxima at 90° and 270° when the field is about 0.5 T. However, it shows minima at the same angles at higher fields. While the latter behavior can be well explained with the SMR scenario, i.e. the $R_{xx}$ is minimum when $M//\sigma$, the anomalous behavior at 0.5 T can be ascribed to the SF. This behavior disappears at a higher temperature (T=60 K), as shown in Figure 4(b), where a larger SMR effect is observed, indicating possibly a higher interfacial ferromagnetic magnetization, coming from the top $CrPS_4$ layer in contact with the heavy metal. This is in good agreement with the larger $R_{xy}$ that we observe at a higher temperature.

Figure 4(c) and 4(d) show the field dependence of $\delta R_{xx}(\gamma)$ at 30 K and 60 K, respectively. Here, $\gamma$ is defined as the angle between the field and x axis when field rotates in the x-z plane, which is zero when the field is along the z axis. Since $M$ is always perpendicular to $\sigma$ in this rotation plane, the magnetoresistance should be a result of only AMR. It is found that no obvious AMR was visible at 60 K, however, a weak AMR signal was observed at 0.5 T and 30 K. This can be ascribed to the AMR effect of the $CrPS_4$ due to the electronic band structure modification by the magnetic field, which is more pronounced at temperatures below $T_N$.

In conclusion, we have determined the magnetic transport properties of the $CrPS_4$/(Pd, Pt) heterostructures. The SF can be unambiguously detected in both systems, indicating the spin-transparent interfaces are formed in those vdW/heavy metal heterostructures. Furthermore, a major enhancement of the magnetic ordering temperature up to room temperature is observed in the Hall effect due to the large spin-orbit coupling at the interface, which suggests potentially a route to achieve room temperature 2D magnetic ordering using a proximity effect. A negative magnetoresistance observed below $T_N$ is explained by the magnetic field modified electronic band structure of the $CrPS_4$. The total magnetoresistance is a superposition of the SMR and the magnetoresistance from

CrPS$_4$. Our results realize the electrical detection of two-dimensional magnetism and provide a new route to realize two-dimensional magnetism above room temperature, highlighting the potential of vdW magnets for spintronic device applications.

**Acknowledgements**

We acknowledge funding from Deutsche Forschungsgemeinschaft (DFG, German Research Foundation) – Spin+X TRR 173 – 268565370 (Projects No. A01 and No. B02), DFG Project No. 358671374, Graduate School of Excellence Materials Science in Mainz (MAINZ) DFG 266, the MaHoJeRo (DAAD Spintronics network, Projects No.57334897 and No. 57524834), the Research Council of Norway (QuSpin Center 262633), National Key Research and Development Program of China (Grants No. 2016YFB0700901, No. 2017YFA0206303, and No. 2017YFA0403701), and National Natural Science Foundation of China (Grants No. 51731001, No. 11675006, No.11805006, No. 11975035, and No. 211012180393).

**References**

1. Jungwirth, T.; Wunderlich, J.; Olejník, K., Spin Hall effect devices. *Nature Materials* **2012,** *11* (5), 382-390.

2. Fan, Y.; Upadhyaya, P.; Kou, X.; Lang, M.; Takei, S.; Wang, Z.; Tang, J.; He, L.; Chang, L.-T.; Montazeri, M., Magnetization switching through giant spin–orbit torque in a magnetically doped topological insulator heterostructure. *Nature Materials* **2014,** *13* (7), 699-704.

3. Han, J.; Richardella, A.; Siddiqui, S. A.; Finley, J.; Samarth, N.; Liu, L., Room-temperature spin-orbit torque switching induced by a topological insulator. *Physical Review Letters* **2017,** *119* (7), 077702.

4. Han, J.; Richardella, A.; Siddiqui, S. A.; Finley, J.; Samarth, N.; Liu, L., Room-Temperature Spin-Orbit Torque Switching Induced by a Topological Insulator. *Physical Review Letters* **2017,** *119* (7), 077702.

5. Sinova, J.; Valenzuela, S. O.; Wunderlich, J.; Back, C. H.; Jungwirth, T., Spin Hall effects. *Reviews of Modern Physics* **2015,** *87* (4), 1213-1260.

6. Nakayama, H.; Althammer, M.; Chen, Y. T.; Uchida, K.; Kajiwara, Y.; Kikuchi, D.; Ohtani, T.; Geprags, S.; Opel, M.; Takahashi, S.; Gross, R.; Bauer, G. E.; Goennenwein, S. T.; Saitoh, E., Spin Hall magnetoresistance induced by a nonequilibrium proximity effect. *Physical Review Letters* **2013,** *110* (20), 206601.

7. Burch, K. S.; Mandrus, D.; Park, J. G., Magnetism in two-dimensional van der Waals materials. *Nature* **2018,** *563* (7729), 47-52.

8. Deng, Y.; Yu, Y.; Song, Y.; Zhang, J.; Wang, N. Z.; Sun, Z.; Yi, Y.; Wu, Y. Z.; Wu, S.; Zhu, J.; Wang, J.; Chen, X. H.; Zhang, Y., Gate-tunable room-temperature ferromagnetism in two-dimensional Fe3GeTe2. *Nature* **2018,** *563* (7729), 94-99.

9. Song, T.; Fei, Z.; Yankowitz, M.; Lin, Z.; Jiang, Q.; Hwangbo, K.; Zhang, Q.; Sun, B.; Taniguchi, T.; Watanabe, K.; McGuire, M. A.; Graf, D.; Cao, T.; Chu, J.-H.; Cobden, D. H.; Dean, C. R.; Xiao, D.; Xu, X., Switching 2D magnetic states via pressure tuning of layer stacking. *Nature Materials* **2019,** *18* (12), 1298-1302.

10. Gong, C.; Li, L.; Li, Z.; Ji, H.; Stern, A.; Xia, Y.; Cao, T.; Bao, W.; Wang, C.; Wang, Y.; Qiu, Z. Q.; Cava, R. J.; Louie, S. G.; Xia, J.; Zhang, X., Discovery of intrinsic ferromagnetism in two-dimensional van der Waals crystals. *Nature* **2017,** *546* (7657), 265-269.

11. Ding, S.; Liang, Z.; Yang, J.; Yun, C.; Zhang, P.; Li, Z.; Xue, M.; Liu, Z.; Tian, G.; Liu, F.; Wang, W.; Yang, W.; Yang, J., Ferromagnetism in two-dimensional $Fe_3GeTe_2$; Tunability by hydrostatic pressure. *Physical Review B* **2021,** *103* (9), 094429.

12. Lohmann, M.; Su, T.; Niu, B.; Hou, Y.; Alghamdi, M.; Aldosary, M.; Xing, W.; Zhong, J.; Jia, S.; Han, W.; Wu, R.; Cui, Y.-T.; Shi, J., Probing Magnetism in Insulating $Cr_2Ge_2Te_6$ by Induced Anomalous Hall Effect in Pt. *Nano Letters* **2019,** *19* (4), 2397-2403.

13. Dolui, K.; Petrović, M. D.; Zollner, K.; Plecháč, P.; Fabian, J.; Nikolić, B. K., Proximity Spin–Orbit Torque on a Two-Dimensional Magnet within van der Waals Heterostructure: Current-Driven Antiferromagnet-to-Ferromagnet Reversible Nonequilibrium Phase Transition in Bilayer $CrI_3$. *Nano Letters* **2020,** *20* (4), 2288-2295.

14. Huang, B.; Clark, G.; Navarro-Moratalla, E.; Klein, D. R.; Cheng, R.; Seyler, K. L.; Zhong, D.; Schmidgall, E.; McGuire, M. A.; Cobden, D. H.; Yao, W.; Xiao,

D.; Jarillo-Herrero, P.; Xu, X., Layer-dependent ferromagnetism in a van der Waals crystal down to the monolayer limit. *Nature* **2017,** *546* (7657), 270-273.

15. Mermin, N. D.; Wagner, H., Absence of Ferromagnetism or Antiferromagnetism in One- or Two-Dimensional Isotropic Heisenberg Models. *Physical Review Letters* **1966,** *17* (22), 1133-1136.

16. Thiel, L.; Wang, Z.; Tschudin, M. A.; Rohner, D.; Gutiérrez-Lezama, I.; Ubrig, N.; Gibertini, M.; Giannini, E.; Morpurgo, A. F.; Maletinsky, P., Probing magnetism in 2D materials at the nanoscale with single-spin microscopy. *Science* **2019,** *364* (6444), 973-976.

17. Jiang, S.; Shan, J.; Mak, K. F., Electric-field switching of two-dimensional van der Waals magnets. *Nature Materials* **2018,** *17* (5), 406-410.

18. Song, T.; Cai, X.; Tu, M. W.; Zhang, X.; Huang, B.; Wilson, N. P.; Seyler, K. L.; Zhu, L.; Taniguchi, T.; Watanabe, K.; McGuire, M. A.; Cobden, D. H.; Xiao, D.; Yao, W.; Xu, X., Giant tunneling magnetoresistance in spin-filter van der Waals heterostructures. *Science* **2018,** *360* (6394), 1214-1218.

19. Song, T.; Tu, M. W.-Y.; Carnahan, C.; Cai, X.; Taniguchi, T.; Watanabe, K.; McGuire, M. A.; Cobden, D. H.; Xiao, D.; Yao, W.; Xu, X., Voltage Control of a van der Waals Spin-Filter Magnetic Tunnel Junction. *Nano Letters* **2019,** *19* (2), 915-920.

20. Calder, S.; Haglund, A. V.; Liu, Y.; Pajerowski, D. M.; Cao, H. B.; Williams, T. J.; Garlea, V. O.; Mandrus, D., Magnetic structure and exchange interactions in the layered semiconductor $CrPS_4$. *Physical Review B* **2020,** *102* (2).

21. Peng, Y.; Ding, S.; Cheng, M.; Hu, Q.; Yang, J.; Wang, F.; Xue, M.; Liu, Z.; Lin, Z.; Avdeev, M.; Hou, Y.; Yang, W.; Zheng, Y.; Yang, J., Magnetic Structure and Metamagnetic Transitions in the van der Waals Antiferromagnet $CrPS_4$. *Advanced Materials* **2020,** *32* (28), e2001200.

22. Zhuang, H. L.; Zhou, J., Density functional theory study of bulk and single-layer magnetic semiconductor $CrPS_4$. *Physical Review B* **2016,** *94* (19).

23. Son, J.; Son, S.; Park, P.; Kim, M.; Tao, Z.; Oh, J.; Lee, T.; Lee, S.; Kim, J.; Zhang, K.; Cho, K.; Kamiyama, T.; Lee, J. H.; Mak, K. F.; Shan, J.; Kim, M.; Park, J. G.; Lee, J., Air-Stable and Layer-Dependent Ferromagnetism in Atomically Thin van

der Waals CrPS$_4$. *ACS Nano* **2021,** *15* (10), 16904-16912.

24. Pei, Q. L.; Luo, X.; Lin, G. T.; Song, J. Y.; Hu, L.; Zou, Y. M.; Yu, L.; Tong, W.; Song, W. H.; Lu, W. J.; Sun, Y. P., Spin dynamics, electronic, and thermal transport properties of two-dimensional CrPS$_4$ single crystal. *Journal of Applied Physics* **2016,** *119* (4), 043902.

25. Tserkovnyak, Y.; Kläui, M., Exploiting Coherence in Nonlinear Spin-Superfluid Transport. *Physical Review Letters* **2017,** *119* (18), 187705.

26. Qaiumzadeh, A.; Skarsvåg, H.; Holmqvist, C.; Brataas, A., Spin Superfluidity in Biaxial Antiferromagnetic Insulators. *Physical Review Letters* **2017,** *118* (13), 137201.

27. Lebrun, R.; Ross, A.; Bender, S. A.; Qaiumzadeh, A.; Baldrati, L.; Cramer, J.; Brataas, A.; Duine, R. A.; Klaui, M., Tunable long-distance spin transport in a crystalline antiferromagnetic iron oxide. *Nature* **2018,** *561* (7722), 222-225.

28. Chen, G.; Qi, S.; Liu, J.; Chen, D.; Wang, J.; Yan, S.; Zhang, Y.; Cao, S.; Lu, M.; Tian, S.; Chen, K.; Yu, P.; Liu, Z.; Xie, X. C.; Xiao, J.; Shindou, R.; Chen, J. H., Electrically switchable van der Waals magnon valves. *Nature Communications* **2021,** *12* (1), 6279.

29. Xing, W.; Qiu, L.; Wang, X.; Yao, Y.; Ma, Y.; Cai, R.; Jia, S.; Xie, X. C.; Han, W., Magnon Transport in Quasi-Two-Dimensional van der Waals Antiferromagnets. *Physical Review X* **2019,** *9* (1).

30. Machado, F. L. A.; Ribeiro, P. R. T.; Holanda, J.; Rodríguez-Suárez, R. L.; Azevedo, A.; Rezende, S. M., Spin-flop transition in the easy-plane antiferromagnet nickel oxide. *Physical Review B* **2017,** *95* (10), 104418.

31. Seki, S.; Ideue, T.; Kubota, M.; Kozuka, Y.; Takagi, R.; Nakamura, M.; Kaneko, Y.; Kawasaki, M.; Tokura, Y., Thermal Generation of Spin Current in an Antiferromagnet. *Physical Review Letters* **2015,** *115* (26), 266601.

32. Lee, H. J.; Helgren, E.; Hellman, F., Gate-controlled magnetic properties of the magnetic semiconductor (Zn,Co)O. *Applied Physics Letters* **2009,** *94* (21), 212106.

33. Wei, D. H.; Niimi, Y.; Gu, B.; Ziman, T.; Maekawa, S.; Otani, Y., The spin Hall effect as a probe of nonlinear spin fluctuations. *Nature Communications* **2012,** *3* (1), 1058.

34. Yang, H.;  Thiaville, A.;  Rohart, S.;  Fert, A.; Chshiev, M., Anatomy of Dzyaloshinskii-Moriya Interaction at Co/Pt Interfaces. *Physical Review Letters* **2015,** *115* (26), 267210.

35. Zhu, W.;  Song, C.;  Han, L.;  Bai, H.;  Wang, Q.;  Yin, S.;  Huang, L.;  Chen, T.; Pan, F., Interface-Enhanced Ferromagnetism with Long-Distance Effect in van der Waals Semiconductor. *Advanced Functional Materials* **2021**.

36. Gu, D.;  Dey, S. K.; Majhi, P., Effective work function of Pt, Pd, and Re on atomic layer deposited $HfO_2$. *Applied Physics Letters* **2006,** *89* (8).

37. Bruno, P.;  Dugaev, V. K.; Taillefumier, M., Topological Hall effect and Berry phase in magnetic nanostructures. *Physical review letters* **2004,** *93* (9), 096806.

38. Swekis, P.;  Markou, A.;  Kriegner, D.;  Gayles, J.;  Schlitz, R.;  Schnelle, W.;  Goennenwein, S. T. B.; Felser, C., Topological Hall effect in thin films of $Mn_{1.5}PtSn$. *Physical Review Materials* **2019,** *3* (1), 013001.

39. Kanazawa, N.;  Onose, Y.;  Arima, T.;  Okuyama, D.;  Ohoyama, K.;  Wakimoto, S.;  Kakurai, K.;  Ishiwata, S.; Tokura, Y., Large Topological Hall Effect in a Short-Period Helimagnet MnGe. *Physical Review Letters* **2011,** *106* (15), 156603.

40. Kurumaji, T.;  Nakajima, T.;  Hirschberger, M.;  Kikkawa, A.;  Yamasaki, Y.;  Sagayama, H.;  Nakao, H.;  Taguchi, Y.;  Arima, T.-h.; Tokura, Y., Skyrmion lattice with a giant topological Hall effect in a frustrated triangular-lattice magnet. *Science* **2019,** *365* (6456), 914.

41. Shao, Q.;  Liu, Y.;  Yu, G.;  Kim, S. K.;  Che, X.;  Tang, C.;  He, Q. L.;  Tserkovnyak, Y.;  Shi, J.; Wang, K. L., Topological Hall effect at above room temperature in heterostructures composed of a magnetic insulator and a heavy metal. *Nature Electronics* **2019,** *2* (5), 182-186.

42. Cheng, Y.;  Yu, S.;  Zhu, M.;  Hwang, J.; Yang, F., Evidence of the Topological Hall Effect in Pt/Antiferromagnetic Insulator Bilayers. *Physical Review Letters* **2019,** *123* (23), 237206.

43. Park, Y.-K.;  Kim, D.-Y.;  Kim, J.-S.;  Nam, Y.-S.;  Park, M.-H.;  Choi, H.-C.;  Min, B.-C.; Choe, S.-B., Experimental observation of the correlation between the interfacial Dzyaloshinskii–Moriya interaction and work function in metallic magnetic trilayers.

*NPG Asia Materials* **2018,** *10* (10), 995-1001.

44. Baldrati, L.; Schmitt, C.; Gomonay, O.; Lebrun, R.; Ramos, R.; Saitoh, E.; Sinova, J.; Klaui, M., Efficient Spin Torques in Antiferromagnetic CoO/Pt Quantified by Comparing Field- and Current-Induced Switching. *Physical Review Letters* **2020,** *125* (7), 077201.

45. Su, T.; Lohmann, M.; Li, J.; Xu, Y.; Niu, B.; Alghamdi, M.; Zhou, H.; Cui, Y.; Cheng, R.; Taniguchi, T.; Watanabe, K.; Shi, J., Current-induced CrI3 surface spin-flop transition probed by proximity magnetoresistance in Pt. *2D Materials* **2020,** *7* (4), 045006.

46. Ross, A.; Lebrun, R.; Ulloa, C.; Grave, D. A.; Kay, A.; Baldrati, L.; Kronast, F.; Valencia, S.; Rothschild, A.; Kläui, M., Structural sensitivity of the spin Hall magnetoresistance in antiferromagnetic thin films. *Physical Review B* **2020,** *102* (9), 094415.

47. Chen, Y.-T.; Takahashi, S.; Nakayama, H.; Althammer, M.; Goennenwein, S. T. B.; Saitoh, E.; Bauer, G. E. W., Theory of spin Hall magnetoresistance. *Physical Review B* **2013,** *87* (14).

48. Dyakonov, M. I., Magnetoresistance due to Edge Spin Accumulation. *Physical Review Letters* **2007,** *99* (12), 126601.